\documentclass[osajnl,twocolumn,showpacs]{revtex4}

\usepackage{amsfonts,amssymb,amsbsy,bm}
\usepackage{graphicx}

\begin{document}

\title{Characterizing the reflectance of
periodic layered media}

\author{J. J. Monz\'on, T. Yonte,
and L. L. S\'anchez-Soto}

\affiliation{Departamento de \'Optica,
Facultad de F\'{\i}sica,
Universidad Complutense, 28040 Madrid,
Spain}

\begin{abstract}
It has recently been shown that periodic layered
media can reflect strongly for all incident angles
and polarizations in a given frequency range. The
standard treatment gets these band gaps from an
eigenvalue equation for the Bloch factor in an
infinite periodic structure. We argue that such
a procedure may become meaningless when dealing
with structures with not very many periods. We
propose an alternative approach based on a
factorization of the multilayer transfer matrix
in terms of three fundamental matrices of simple
interpretation. We show that the trace of the
transfer matrix sorts the periodic structures
into three types with properties closely
related to one (and only one) of the three
fundamental matrices. We present the reflectance
associated to each one of these types, which
can be considered as universal features of
the reflection in these media.
\end{abstract}

%\ocis{230.4170, 120.5700, 120.7000, 000.3860}

\section{Introduction}

Wave propagation in periodic layered media is
very similar to the motion of electrons in
crystallyne solids (hence, they are also known
as photonic crystals). Thus, it is hardly surprising
that some well-established concepts of solid-state
physics (such as Bloch waves, Brillouin zones, and
forbidden bands) have been incorporated in the
doctrine of photonic crystals~\cite{JO95}.

The standard approach~\cite{YE88} identifies the
periodic structure with a one-dimen\-sional lattice
that is invariant under lattice translations, which
requires the system to be strictly infinite. Of course,
this is unattainable in practice. In solid-state physics
one goes around this difficulty by considering
that the one-dimensional lattice closes into a ring;
i.e., by imposing periodic boundary conditions.
While this seems to be reasonable for a lattice
of atoms, is obviously unfeasible for periodically
stratified media.

In spite of this fact, only under the assumption
of an endlessly repeating stack of dielectric
slabs this standard treatment gets an eigenvalue
equation for the Bloch factor and then one
finds~\cite{YE77,FI98,DO98,YA98,LE00} that strong
reflection (stop bands) occurs when the trace of
the $2 \times 2$ matrix describing the basic
period exceeds 2 in magnitude.

In practice not very many periods (say, ten)
are needed to have a stop band. Therefore,
one is led to consider stacks of $N$ periods,
which are often appropriately called ``finite"
periodic structures~\cite{LE94}. The matrix
describing these structures can be computed as
the $N$th power of the basic period in terms of
Chebyshev  polynomials~\cite{BO99,LE87}. This
gives an exact expression for the reflectance,
but so complicated that it is difficult to get,
at first glance, any feeling about the
performance of the system. Moreover, the
distinction between allowed and forbidden
bands is obtained only numerically.

The aim of this paper is to provide an
alternative setting for dealing with these
finite periodic structures that stresses
the different scaling laws appearing in
the reflectance. To this end we resort to
the Iwasawa decomposition to represent the
action of any multilayer as the product
of three fundamental matrices of simple
interpretation. We show that these matrices
present quite distinct reflectances.
The trace of the basic period allows us
to classify periodic structures in three
types with properties closely related to
one (and only one) of the three fundamental
matrices. Finally, we study the reflectance
associated to each one of these types.

The advantage of this approach is, in our
view, twofold: it naturally emphasizes
the different laws for the forbidden bands,
allowed bands, and band edges and, at the same
time, provides universal expressions
remarkably simple for the corresponding
reflectances.

\section{The origin of the bandgap in infinite
periodic structures}

To keep the discussion as self-contained
as possible, we first briefly summarize
the essential ingredients of multilayer
optics that we shall need for our purposes.
The basic period of the structure is
illustrated in Fig.~1 and consists of a
stack of $1, \ldots, j, \ldots, m$
plane-parallel layers sandwiched between
two semi-infinite ambient ($a$) and
substrate ($s$) media, which we shall
assume to be identical, since this is the
common experimental case. Hereafter all
the media are supposed to be lossless,
homogeneous, and isotropic.

\begin{figure}
\centering
\resizebox{0.60\columnwidth}{!}{\includegraphics{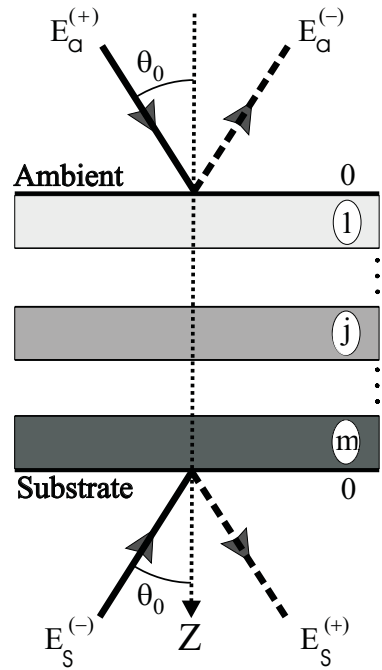}}
\caption{Wave vectors of the input [$E_a^{(+)}$ and $E_s^{(-)}$]
and the output [$E_a^{(-)}$ and $E_s^{(+)}$] fields in a
multilayer sandwiched between two identical semi-infinite ambient
and substrate media.}
\end{figure}

We consider an incident monochromatic linearly polarized plane
wave from the ambient that makes an angle $\theta_0$ with the
normal to the first interface and  has amplitude $E_{a}^{(+)}$.
The electric field is either in the plane of incidence ($p$
polarization) or perpendicular to the plane of incidence ($s$
polarization). We consider as well another plane wave of the same
frequency and polarization, and with amplitude $E_{s}^{(-)}$,
incident from the substrate at the same  angle $\theta _{0}$.

As a result of multiple reflections in all
the interfaces~\cite{AZ87}, we have a
backward-traveling plane wave in the ambient,
denoted $E_{a}^{(-)}$, and a forward-traveling
plane wave in the substrate, denoted $E_{s}^{(+)}$.
If we take the field amplitudes as a vector
of the form
\begin{equation}
\label{Evec} \vec{E} = \left ( \begin{array}{c}
E^{(+)} \\
E^{(-)}
\end{array}
\right )\ ,
\end{equation}
which applies to both ambient and substrate
media, then the amplitudes at each side
of the multilayer are related by the $2 \times 2$
complex matrix $\mathsf{M}_{as}$, we shall
call the transfer matrix, in the form
\begin{equation}
\label{M1} \vec{E}_a = \mathsf{M}_{as} \, \vec{E}_s\ .
\end{equation}
The matrix  $\mathsf{M}_{as}$ can be shown to
be of the form~\cite{YO02,MO02}
\begin{equation}
\label{Mlossless}
\mathsf{M}_{as} =
\left (
\begin{array}{cc}
1/T_{as} & R _{as}^\ast/T_{as}^\ast \\
R_{as}/T_{as} & 1/T_{as}^\ast
\end{array}
\right )
\equiv
\left (
\begin{array}{cc}
\alpha & \beta \\
 \beta^\ast & \alpha^\ast
\end{array}
\right )  ,
\end{equation}
where the complex numbers
\begin{equation}
R_{as}  =  | R_{as} | \exp (i \rho),
\qquad
T_{as}  =  | T_{as} | \exp (i \tau) ,
\end{equation}
are, respectively, the overall reflection and
transmission coefficients for a wave incident
from the ambient. Note that  we have
$ \det \mathsf{M}_{as}= +1$, which is
equivalent to $|R_{as}|^2  + |T_{as}|^2 = 1$,
and then  the set of lossless multilayer
matrices reduces to the group SU(1,1), whose
elements depend on three independent real
parameters~\cite{MO99a,MO99b}.

The identity matrix corresponds to $T_{as} =1$ and
$R_{as} = 0$, so it  represents an antireflection
system without transmission phase shift.  The matrix
that describes the overall system obtained by putting
two multilayers together is the product of the matrices
representing each one of them, taken in the
appropriate order. So, two multilayers, which are
inverse, when composed give an antireflection system.

When we repeat the basic period $\mathsf{M}_{as}$
the fields at the $n$th cell are given by
\begin{equation}
\label{fixed} \vec{E}_{n+1} =
\mathsf{M}_{as} \vec{E}_{n} .
\end{equation}
In an infinite structure the Floquet theorem
ensures that these cells are equivalent and
so the two vectors in Eq.~(\ref{fixed}) are
proportional~\cite{YE88,LE94}
\begin{equation}
\vec{E}_{n+1} = \lambda \vec{E}_{n} ,
\end{equation}
and therefore the Bloch factor $\lambda$ is
determined as the eigenvalue of the basic-period
transfer matrix $\mathsf{M}_{as}$:
\begin{equation}
\label{standard}
(\mathsf{M}_{as} - \lambda ) \vec{E}_n = 0 ,
\end{equation}
which gives
\begin{equation}
\label{eig}
\lambda^2 - [ \mathrm{Tr}(\mathsf{M}_{as}) ]
\lambda  + 1 = 0 .
\end{equation}
If $[\mathrm{Tr}( \mathsf{M}_{as})]^2 > 4$, the
solution grows or decays exponentially and
no propagating wave is possible (stop band).
On the contrary, Bloch waves appear in this
algebraic approach when $[\mathrm{Tr}
( \mathsf{M}_{as})]^2 < 4$ and then the matrix $
\mathsf{M}_{as}$ has two eigenvalues (allowed band).
The band edges are obtained by setting
$[\mathrm{Tr}( \mathsf{M}_{as})]^2 = 4$. Such
a periodic structure exhibits resonance reflection
very much like the diffraction of X rays by
crystal lattice planes and, in consequence, it is
usually called a Bragg reflector~\cite{YE88}.
By properly designing the basic period it
is possible to achieve extremely high
reflectance for some selected spectral
region~\cite{SO99}.

\section{Finite periodic structures and
Iwasawa decomposition}

As anticipated in the Introduction, an infinite
multilayer is unrealistic in practice, and one
should consider $N$-period finite structures,
for which the previous standard approach
fails~\cite{BE96,CO01}.

The overall transfer matrix $\mathsf{M}^{(N)}_{as}$
for a $N$-period structure is equal to $(\mathsf{M}_{as})^N$.
Explicit but complicated formulas for the
corresponding matrix elements exist~\cite{BO99,LE87},
but they do not give analytic expressions for
the bandgap and are difficult to analyze.

We have recently advocated using the Iwasawa
decomposition to study layered media~\cite{MO01b},
since it provides a remarkable factorization of the
matrix representing  any multilayer  (no matter how
complicated it could be)  as the product  of three
matrices of simple interpretation. The decomposition
can be stated as
\begin{equation}
\label{Iwa1}
\mathsf{M}_{as} =
\mathsf{K} (\phi)
\mathsf{A} (\xi)
\mathsf{N} (\nu) ,
\end{equation}
where
\begin{eqnarray}
\label{Iwasa1}
\mathsf{K} (\phi) & = &
\left (
\begin{array}{cc}
\exp (i\phi/2) & 0 \\
0 & \exp (-i\phi/2)
\end{array}
\right ) ,
\nonumber \\
\mathsf{A} (\xi) & = &
\left (
\begin{array}{cc}
\cosh (\xi/2) & i \sinh(\xi/2) \\
-i \sinh(\xi/2) & \cosh (\xi/2)
\end{array}
\right ) , \\
\mathsf{N} (\nu) & = &
 \left (
\begin{array}{cc}
1 - i \nu/2& \nu/2 \\
\nu/2 & 1+ i \nu/2
\end{array}
\right ) .
\nonumber
\end{eqnarray}
The parameters $\phi, \xi$, and $\nu$
are given in terms of the elements of
the multilayer matrix by
\begin{eqnarray}
\label{param}
\phi/2 & = & \arg (\alpha + i \beta)\ ,
\nonumber  \\
\xi/2  &  =  & \ln  (1/|\alpha +i \beta | )\ , \\
\nu/2 & = & \mathrm{Re} (\alpha  \beta^\ast)/
|\alpha + i \beta |^2\ ,  \nonumber
\end{eqnarray}
and their ranges  are  $\xi,  \nu \in \mathbb{R}$
and $-2\pi \le \phi \le 2\pi$.  Now, we can interpret
the physical action of the matrices appearing in
Eq.~(\ref{Iwa1}). $\mathsf{K}(\phi)$ represents the
free propagation of the fields $\vec{E}$ in the
ambient medium through an optical phase thickness
of $\phi/2$, which reduces to a mere shift of the
origin of phases. Alternatively, this can be seen
as an antireflection system. The second matrix
$\mathsf{A}(\xi)$ represents a symmetric
system with real transmission coefficient
$T_{\mathsf{A}} = \mathrm{sech}(\xi/2)$ and
reflection phase shift $\rho_{\mathsf{A}} = \pm \pi/2.$
Finally, the third matrix, $\mathsf{N}(\nu)$, represents
a system having $T_{\mathsf{N}} = \cos(\tau_{\mathsf{N}})
\exp(i \tau_{\mathsf{N}})$  and $R_{\mathsf{N}} = \sin
(\tau_{\mathsf{N}}) \exp(i \tau_{\mathsf{N}})$, with
$\tan (\tau_{\mathsf{N}}) = \nu/2$.
In Ref.~\cite{YO02} we have provided simple examples
of realistic systems that accomplish these requirements.

For our purposes here, the essential point is that
the reflectance $\mathfrak{R} = | R |^2$ associated
to each one of these matrices is
\begin{eqnarray}
\label{RefKAN}
\mathfrak{R}_{\mathsf{K}} & = & 0 , \nonumber \\
\mathfrak{R}_{\mathsf{A}} & = & \tanh^2 (\xi/2) , \\
\mathfrak{R}_{\mathsf{N}} & = &  (\nu/2)^2/[1 + (\nu/2)^2] .
\nonumber
\end{eqnarray}
While $\mathfrak{R}_{\mathsf{K}}$ is identically
zero, $\mathfrak{R}_{\mathsf{A}}$ and
$\mathfrak{R}_{\mathsf{N}}$ tend to unity when
$\xi$ and $\nu$, respectively, increase. However,
they have distinct growth: $\mathfrak{R}_{\mathsf{A}}$
goes to unity exponentially, while
$\mathfrak{R}_{\mathsf{N}}$ goes as
$O(\nu^{-2})$.

The Iwasawa decomposition proves also to be a
powerful tool for the classification of layered
media. Indeed, after Ref.~\cite{SA01} one is led
to introduce the following criterion: a matrix is
of type $\mathsf{K}$ when $[\mathrm{Tr} (
\mathsf{M}_{as})]^2 < 4$, is of type $\mathsf{A}$
when  $[\mathrm{Tr} ( \mathsf{M}_{as})]^2 > 4$, and
finally is of type $\mathsf{N}$ when $[\mathrm{Tr}
(\mathsf{M}_{as})]^2  = 4$.  Although this trace
criterion has an elegant geometrical
interpretation~\cite{MO02} and coincides with the
one giving the stop bands in Eq.~(\ref{eig}),
let us remark that if a multilayer has a transfer
matrix $\mathsf{M}_{as}$ of type $\mathsf{K}$,
$\mathsf{A}$, or $\mathsf{N}$, one can always
find a family of matrices $\mathsf{C}$ [also in SU(1,1)]
such that
\begin{equation}
\label{Mconj}
\widehat{ \mathsf{M}}_{as} = \mathsf{C} \
\mathsf{M}_{as} \ \mathsf{C}^{-1}
\end{equation}
is just a matrix $\mathsf{K}(\phi)$, $\mathsf{A}(\xi)$,
or  $\mathsf{N}(\nu)$, respectively. This
conjugation by $\mathsf{C}$ preserves the trace:
$ \mathrm{Tr} (\widehat{\mathsf{M}}_{as})  =
\mathrm{Tr} ( \mathsf{M}_{as})$. Moreover,
Eq.~(\ref{Mconj}) allows us to recast the
multilayer action in Eq.~(\ref{M1}) as
\begin{equation}
\label{M2} \widehat{\vec{E}}_a =
\widehat{ \mathsf{M}}_{as} \widehat{\vec{E}}_s ,
\end{equation}
where the new field vectors are
\begin{equation}
\widehat{\vec{E}} = \mathsf{C} \vec{E} .
\end{equation}
In other words, the matrix $\mathsf{C}$
gives a new vector basis such that the
basic period of the system, when viewed in
such a basis, presents a reflectance exactly of
the form in Eq.~(\ref{RefKAN}). This provides
a full characterization of the reflectance
of any periodic system, as we shall demonstrate
in next Section.

\section{Characterizing the three fundamental
behaviors of the reflectance}

Since the matrix $\widehat{\mathsf{M}}_{as}$
associated to the basic period belongs to one of
the subgroups $\mathsf{K}(\phi)$, $\mathsf{A} (\xi)$,
or $\mathsf{N}(\nu) $ of the Iwasawa decomposition,
and all these subgroups are, in our special case,
Abelian and uniparametric, we have that
\begin{equation}
\widehat{\mathsf{M}}_{as} (\mu_1)
\widehat{\mathsf{M}}_{as} (\mu_2)
= \widehat{\mathsf{M}}_{as} (\mu_1 + \mu_2) ,
\end{equation}
where $\mu$ represents the appropriate parameter
$\phi$, $\xi$, or $\nu$. For a $N$-period
system the overall transfer matrix $\mathsf{M}^{(N)}_{as}$
is
\begin{equation}
\label{Npow}
\mathsf{M}^{(N)}_{as} =
\mathsf{C}^{-1} \
[ \widehat{\mathsf{M}}_{as} (\mu) ]^N \
\mathsf{C} =
\mathsf{C}^{-1} \
\widehat{\mathsf{M}}_{as} (N \mu) \
\mathsf{C} ,
\end{equation}
which does not depend on the explicit form
of the basic period.

After Eq.~(\ref{Npow}), one must expect three
universal behaviors of the reflectance
according the basic-period transfer matrix
is of the type $\mathsf{K}$, $\mathsf{A}$ or
$\mathsf{N}$. We shall work in what follows the
detailed structure of these three basic laws.

Because the stop bands are given by the condition
$[\mathrm{Tr}(\mathsf{M}_{as})]^2 > 4$, we
first consider the case when $\mathsf{M}_{as}$
is of type $\mathsf{A}$. Then, Eq.~(\ref{Mconj})
gives
\begin{equation}
\label{conjC}
\mathsf{M}_{as} = \mathsf{C}^{-1} \
\mathsf{A}(\chi) \  \mathsf{C} ,
\end{equation}
where we have denoted
\begin{equation}
\mathrm{Re}(\alpha) =
\frac{1}{2} \mathrm{Tr}\
( \mathsf{M}_{as}) =  \cosh (\chi) > 1 ,
\end{equation}
because we are taking into account only
positive values of $\mathrm{Re}(\alpha)$, since
negative values can be treated much in same way.
If we put
\begin{equation}
\mathsf{C} =
\left (
\begin{array}{cc}
\ \mathfrak{c}_1 \ & \ \mathfrak{c}_2 \ \\
\ \mathfrak{c}^\ast_2 \ & \ \mathfrak{c}^\ast_1 \
\end{array}
\right ) \ ,
\end{equation}
one solution of Eq.~(\ref{conjC}) is
\begin{equation}
\mathfrak{c}_1 = F (\beta^\ast + i \ \sinh \chi ) ,
\qquad
\mathfrak{c}_2 = - i  F  \ \mathrm{Im}(\alpha),
\end{equation}
where the factor $F$ is
\begin{equation}
F = \frac{1}
{\sqrt{2 \sinh \chi [ \sinh \chi - \mathrm{Im}(\beta)]}} .
\end{equation}

Carrying out the matrix multiplications in
Eq.~(\ref{Npow}) it is easy to obtain the reflectance
of the $N$-period system as
\begin{equation}
\label{RasA}
\mathfrak{R}_{\mathsf{A}}^{(N)} =
\frac{ | \beta|^2}{| \beta|^2 +
[\sinh ( \chi) / \sinh (N \chi) ]^2 } .
\end{equation}
This is an exact expression for any value of $N$.
As $N$ grows, $\mathfrak{R}_{\mathsf{A}}^{(N)}$
approaches unity exponentially, as one could expect
from a band stop.

The band edges are determined by the limit condition
$[\mathrm{Tr}(\mathsf{M}_{as})]^2 = 4$; that is,
when $\mathsf{M}_{as}$ is of type $\mathsf{N}$.
A calculation very similar to the previous one shows
that now
\begin{equation}
\label{RasN}
\mathfrak{R}_{\mathsf{N}}^{(N)}  = \frac{ | \beta|^2}
{| \beta|^2 + ( 1/ N ) ^2 } ,
\end{equation}
with a typical behavior $\mathfrak{R}_{\mathsf{N}}^{(N)}
\sim 1 - O(N^{-2})$ that is universal in the physics
of reflection. The general results (\ref{RasA})
and  (\ref{RasN}) have been obtained in a different
framework by Yeh~\cite{YE88} and Lekner~\cite{LE94}.

Finally, in the allowed bands we have $[\mathrm{Tr}
(\mathsf{M}_{as})]^2 < 4$; $\mathsf{M}_{as}$ is of
type $\mathsf{K}$, and
\begin{equation}
\label{RasK}
\mathfrak{R}_{\mathsf{K}}^{(N)}  =
\frac{ \mathcal{Q}^2 - 2  \mathcal{Q}
\cos (2 N \Omega )}{1 + \mathcal{Q}^2 -
 2  \mathcal{Q} \cos (2 N \Omega )} ,
\end{equation}
where
\begin{equation}
\mathcal{Q} = \frac{ |\beta |^2}
{ |\beta |^2 - | \alpha - e^{i \Omega} |^2} ,
\end{equation}
and the phase $\Omega$ is determined by
\begin{equation}
e^{i \Omega} = (\mathrm{Re} \ \alpha)
+ i \sqrt{1 - (\mathrm{Re}\ \alpha)^2} .
\end{equation}
Now the reflectance oscillates with $N$
between the values $(\mathcal{Q}^2 -
2\mathcal{Q})/(\mathcal{Q} - 1)^2$ and
$(\mathcal{Q}^2 + 2\mathcal{Q})/
(\mathcal{Q} + 1)^2$.

Equations (\ref{RasA}), (\ref{RasN}) and
(\ref{RasK}) are the three fundamental
reflectances we were looking for. In our
opinion, they have the virtue of simplicity.
In practice, the basic period usually
consists of two homogeneous layers and
the expressions become simpler~\cite{BO99}.
We emphasize that the scaling laws
expressed by these equations are
universal features of the reflection
in finite periodic systems.

\section{Conclusions}

We have used a trace criterion to provide
in a direct way three fundamental forms for
the reflectance that are universal for any
multilayer. When the system is finite
periodic this result has allowed us to
obtain in a direct way explicit expressions
for the reflectance of the overall system.

This approach has physical meaning on its own, irrespective
whether the system is periodic or not. Moreover, it leads
naturally to a further analysis of finite periodic
bandgap structures that does not need the strict requirement
of infinite periodicity, as it is assumed in the standard
theory.

We acknowledge illuminating discussions with
Jos\'e F. Cari\~{n}ena.

\end{document}